\documentclass[onecolumn,preprintnumbers,amsmath,amssymb,nofootinbib,showpacs]{revtex4}

\usepackage{epsf}
\usepackage{graphicx}
\usepackage{bm}
\usepackage{amsmath}

\newcommand{\be}{\begin{equation}}
\newcommand{\ee}{\end{equation}}
\newcommand{\bea}{\begin{eqnarray}}
\newcommand{\eea}{\end{eqnarray}}
\newcommand{\bdm}{\begin{displaymath}}
\newcommand{\edm}{\end{displaymath}}
\newcommand{\beas}{\begin{eqnarray*}}
\newcommand{\eeas}{\end{eqnarray*}}
\newcommand{\av}[1]{\left< #1\right>}
\newcommand{\intdom}[1]{\frac{1}{V}\int_{\mathcal{D}} #1\sqrt{h}d^3\mathbf{x}}
\newcommand{\dom}[1]{#1_{\mathcal{D}}}
\newcommand{\Qd}{\mathcal{Q}_{\mathcal{D}}}
\newcommand{\Rd}{\mathcal{R}_{\mathcal{D}}}
\newcommand{\Pd}{\mathcal{P}_{\mathcal{D}}}
\newcommand{\Td}{\mathcal{T}_{\mathcal{D}}}
\newcommand{\bkr}{\overline{\rho}}

\begin{document}

\preprint{HD-THEP-07-32}

\title{Cosmological Backreaction from Perturbations}

\author{Juliane Behrend}
\email{J.Behrend@thphys.uni-heidelberg.de}
\author{Iain A. Brown}
\email{I.Brown@thphys.uni-heidelberg.de}
\author{Georg Robbers}
\email{G.Robbers@thphys.uni-heidelberg.de}
\affiliation{Institut f\"ur Theoretische Physik, Philosophenweg 16, 69120 Heidelberg, Germany}

\date{\today}

\begin{abstract}
We write the averaged Einstein equations in a form suitable for use with Newtonian gauge linear perturbation theory and track the size of the modifications to standard Robertson-Walker evolution on the largest scales as a function of redshift for both Einstein de-Sitter and $\Lambda$CDM cosmologies. In both cases the effective energy density arising from linear perturbations is of the order of $10^{-5}$ times the matter density, as would be expected, with an effective equation of state $w_\mathrm{eff}\approx -1/19$. Employing a modified Halofit code to extend our results to quasilinear scales, we find that, while larger, the deviations from Robertson-Walker behaviour remain of the order of $10^{-5}$.
\end{abstract}


\maketitle

\section{Introduction}
\label{intro}
Observations of the cosmic microwave background (CMB) \cite{WMAP,WMAP(2),WMAP3}, large-scale structure (LSS) (e.g. \cite{2df,SDSS}) and supernovae type Ia data \cite{RiessEtAl98,PerlmutterEtAl98,RiessEtAl06,WoodVaseyEtAl07} consistently indicate the presence in the universe of significant ``dark'' components. Data from the LSS and from nucleosynthesis indicates that the density of baryonic matter should not exceed about 5\% of the critical density, while data from the CMB suggests that the universe should be flat, implying that around 95\% of the matter-energy content of the universe is unobserved. In combination with numerical studies (e.g. \cite{SpringelEtAl05}) and observations of the baryonic acoustic oscillations in the LSS \cite{2df,SDSS} this has lead to the ``concordance model'' in which roughly 5\% of the universe is in the form of baryonic matter, 20\% of the universe is in the form of some ``dark matter'' that interacts only gravitationally, and about 75\% is in the form of a ``dark energy'' fluid with a negative equation of state today $w\approx-1$.

The most popular alternatives to a cosmological constant involve exotic fluids that violate the strong energy condition in the late universe, of which scalar-field models such as quintessence are the most popular (e.g. \cite{Wetterich88,RatraPeebles88,CaldwellEtAl97} for pioneering works and \cite{CopelandEtAl06,Linder07} for recent reviews.) There are also many other modifications of both the matter and gravitational sectors. Another approach is to consider the impact of local inhomogeneities on the luminosity distances in the local universe (see for example \cite{EnqvistMattson07,Kasai07,Marra07} for some recent work on this topic). The modifications that the local inhomogeneities introduce can, it has been suggested, account for many of the features exhibited by dark energy.

However, the assumptions used to build up our standard model include an implicit averaging procedure, assuming that the averaged Einstein tensor is equivalent to the Einstein tensor built from an average metric -- and such is not the case \cite{Ellis84,Futamase89}. The Einstein equations are non-linear and local; the correct approach is to average the local equations across some domain rather than to assume ab initio ``averaged'' equations. Studies into averaged cosmologies include \cite{Kasai93,Futamase96,BuchertEhlers95,Boersma97,RussEtAl97} and the impact of inhomogeneities were applied to account for a dark energy in \cite{Buchert99}. In the years since, this ``backreaction'' and the impact of inhomogeneities on observables has been relatively well studied in a variety of models and remains an active field \cite{Buchert01,Wetterich01,BuchertCarfora02,Rasanen03,Rasanen04,KolbEtAl04,EllisBuchert05,AlnesEtAl05,Rasanen06,IshibashiWald06,BuchertEtAl06,LarenaBuchert06,LarenaEtAl06,BiswasEtAl06,KasaiEtAl06,VanderveldEtAl06,TanakaEtAl06,Brouzakis06,LiSchwarz07,Enqvist07,Wiltshire07,Mattsson07,Ishak07,VanderveldEtAl07,Wiltshire07_2,KhosraviEtAl07,Hossain07}; see also \cite{BuchertReview07} for a recent status report. The attraction of this approach is that it can recover quantities that -- in principle -- could act as a dark energy, without the need for exotic matter components or modifications to general relativity. Moreover, one can intuitively state that as the universe grows increasingly non-linear, the deviations on averaging from a truly homogeneous and isotropic model should increase, potentially providing an appealing solution to the coincidence problem. The current universe is significantly inhomogeneous at scales of up to 100-300Mpc/$h$ \cite{PercivalEtAl06}. While we might still expect to recover an FLRW-like metric when averaging across scales larger than 100Mpc/$h$ (see \cite{Lu07,McClure07} for recent studies) there is no automatic guarantee that it will obey the familiar Friedmann and Raychaudhuri equations. It has been shown in many studies that structure -- both at the linear order and for highly inhomogeneous models -- can introduce modifications to the large scale effective Friedmann equations (though this is still somewhat debated; e.g. \cite{AlnesEtAl05_2,KasaiEtAl06,TanakaEtAl06,EnqvistMattson07}). Even should the effect be relatively minor, with current and future experiments a divergence from Robertson-Walker behaviour on the order of $10^{-3}$ could be significant.

In this paper we calculate numerically as a function of redshift the size of the deviations from standard Robertson-Walker behaviour arising from linear and quasilinear perturbations. Such a calculation complements the recent studies by \cite{VanderveldEtAl07}, in which the authors reconstruct the impact of backreaction effects from the observational data, and \cite{KhosraviEtAl07} wherein the authors evaluate the size of the effective density of the backreaction as a function of redshift in a structured Robertson-Walker model. Much of the current literature concerns exact inhomogeneous rather than perturbative models, not least because the impact in a perturbative approach is not expected to be large. However, the magnitude and nature of the backreaction and other corrections associated with averaging that arise naturally within a standard Einstein de-Sitter or $\Lambda$CDM cosmology remain open questions and this should be addressed. We work in conformal Newtonian gauge, in contrast to much of the literature which concerns itself with synchronous gauge, for two main reasons. Firstly Newtonian gauge, unlike synchronous gauge, is well-defined and should yield unambiguous conclusions; and, moreover, the use of Newtonian gauge enables us to employ the easily extensible cmbeasy \cite{Doran03} code to track the size of the perturbations. We find that linear perturbations for a model Einstein de-Sitter universe introduce an effective energy density $\rho_\mathrm{eff}\approx (4\times 10^{-5})\bkr_m$ with an equation of state $p_\mathrm{eff}/\rho_\mathrm{eff}=w_\mathrm{eff}\approx -1/19$, while those for the $\Lambda$CDM concordance model introduce $\rho_\mathrm{eff}\approx (1.3\times 10^{-5})\bkr_m$ with a marginally lower equation of state. Employing a modified Halofit \cite{SmithEtAl02} code to include quasilinear scales increases the effective energy densities to $\rho_\mathrm{eff}\approx (5.6\times 10^{-5})\bkr_m$ and $\rho_\mathrm{eff}\approx(1.6\times 10^{-5})\bkr_m$ for the EdS and $\Lambda$CDM cases respectively, with unaltered equations of state. These are in good agreement with other evaluations of $\rho_\mathrm{eff}$ (e.g. \cite{Wetterich01,Rasanen04,VanderveldEtAl07,KhosraviEtAl07}) which typically remain of the same order.

We begin with a brief overview of the 3+1 formalism of general relativity and in \S\ref{Averaging} briefly discuss averages in cosmology before applying a simple average in \S\ref{Averages}. In \S\ref{NewtGauge} we calculate the forms of the backreaction terms for conformal Newtonian gauge and in \S\ref{Backfast} we present the deviations from an FLRW model from linear perturbations employing a modified Boltzmann code for both Einstein de-Sitter and $\Lambda$CDM universes, making it more general than the similar studies in \cite{Rasanen04,KasaiEtAl06}, who did not consider $\Lambda$CDM universes, and \cite{TanakaEtAl06}, who worked in a rather narrow range of domain sizes. We follow by estimating the impact from quasilinear scales in \S\ref{Halofit}.

\section{Einstein Equations in 3+1 form: Basic Quantities and Results}
\label{ADM}
We here briefly summarise the main results of the 3+1 formalism of general relativity and refer the readers to \cite{Gourgoulhon07,York79,Wald,Buchert01} for details. Provided that the spacetime $(\mathcal{M},\bf{g})$ is globally hyperbolic we may foliate $\mathcal{M}$ with a family of spacelike hypersurfaces $\Sigma_t$, such that each hypersurface is a level surface of a smooth scalar field $t$, which we will later identify with the time coordinate. The normal unit vector $\bf{n}$ to the surface $\Sigma_t$ being timelike must satisfy
\be
  \mathbf{n}\cdot\mathbf{n}=-1 .
\ee
It is necessarily colinear to the metric dual $\boldsymbol{\nabla}t$ of the gradient 1-form $\mathbf{d}t$ and the proportionality factor is called the lapse function $\alpha$
\be
  \mathbf{n}=-\alpha\boldsymbol{\nabla}t ,
\ee
where $\alpha\equiv\left(-\boldsymbol{\nabla}t\cdot\boldsymbol{\nabla}t\right)^{-1/2}=-\left(\langle\mathbf{d}t,\boldsymbol{\nabla}t\rangle\right)^{-1/2}$. To ensure that $\mathbf{n}$ is future oriented, $\alpha$ must be larger than zero and in particular never vanishes for a regular foliation. Since $\langle\mathbf{d}t,\alpha\mathbf{n}\rangle=1$ the hypersurfaces are Lie dragged by the normal evolution vector $\alpha\mathbf{n}$.

Now introduce coordinate systems $(x^i)$ on each hypersurface $\Sigma_t$ that vary smoothly between neighbouring hypersurfaces, such that $(x^\mu)=(t,x^1,x^2,x^3)$ form well-behaved coordinate systems on $\mathcal{M}$. The 1-form $\mathbf{d}t$ is dual to the time vector $\boldsymbol{\partial}_t$. Hence, just like the normal evolution vector, the time vector Lie drags the hypersurfaces. In general these vectors do not coincide. The difference defines the shift vector $\boldsymbol{\beta}$
\be
  \boldsymbol{\partial}_t=\alpha\mathbf{n}+\boldsymbol{\beta} .
\label{shift}
\ee
It is purely spatial since $\langle\mathbf{d}t,\boldsymbol{\beta}\rangle=\langle\mathbf{d}t,\boldsymbol{\partial}_t\rangle-\langle\mathbf{d}t,\alpha\mathbf{n}\rangle=0$. In terms of the natural basis $(\partial_\mu)$ of the introduced coordinates $(x^\mu)$ the normal unit vector can then be expressed as
\be
  n^\mu=\frac{1}{\alpha}(1,-\beta^i) .
\ee 
With the projection operator into $\Sigma_t$
\be
  h_{\mu\nu}=g_{\mu\nu}+n_\mu n_\nu 
\ee
the components of the 3-metric $h_{ij}$ become
\be
  h_{ij}=g_{\mu\nu}h^\mu_i h^\nu_j .
\ee
From here one may find that the line element in these coordinates is
\be
  ds^2=(-\alpha^2+\beta_i\beta^i)dt^2+2\beta_idtdx^i+h_{ij}dx^idx^j ,
\ee
where $\beta_i=h_{ij}\beta^j$.

The projection of the gradient $\nabla_\mu$ of the unit normal vector defines the extrinsic curvature tensor
\be
  K_{ij}=-h_i^k\nabla_kn_j .
\ee
Using Frobenius' theorem this can be expressed as the Lie derivative 
\be
  K_{ij}=-\frac{1}{2}\mathcal{L}_nh_{ij}
\label{excurv}
\ee
and the evolution of the 3-metric $h_{ij}$ is given by the Lie derivative along the normal evolution vector $\alpha\mathbf{n}$. Using (\ref{shift}) and (\ref{excurv}) the evolution equation becomes
\be
  \alpha\mathcal{L}_nh_{ij}=\mathcal{L}_{\partial_t}h_{ij}-\mathcal{L}_\beta h_{ij}=\dot{h}_{ij}-2\mathcal{D}_{(i}\beta_{j)}=-2\alpha K_{ij}
\ee
where an overdot denotes $\partial/\partial t$, $\mathcal{D}_i$ is the covariant derivative on the 3-surface and brackets denote symmetrisation on the enclosed indices.

Now we perform a 3+1 decomposition of the stress-energy tensor
\be
  T_{\mu\nu}=\rho n_\mu n_\nu+2n_{(\mu}j_{\nu)}+S_{\mu\nu} \iff
  \rho=T_{\mu\nu}n^\mu n^\nu, \quad j_i=-n^\mu T_{i\mu}, \quad S_{ij}=T_{ij}
\ee
and project the Einstein equations onto the hypersurface and along its normal. This results in the Hamilton constraint equation
\be
  \mathcal{R}+K^2-K^i_jK^j_i=16\pi G\rho+2\Lambda,
\ee
where $\mathcal{R}$ is the Ricci scalar on $\Sigma_t$ and $K=K^i_i$; the momentum constraint equation,
\be
  \mathcal{D}_j\left(K^{ij}-h^{ij}K\right)=8\pi Gj^i ;
\ee
and the evolution equation
\bea
  \dot{K}_{ij}&=&\alpha\left(\mathcal{R}_{ij}-2K^a_iK_{aj}+KK_{ij}-8\pi GS_{ij}+4\pi Gh_{ij}\left(S-\rho\right)-\Lambda h_{ij}\right) \nonumber \\
    && \quad -\frac{1}{\alpha}\left(\mathcal{D}_i\mathcal{D}_j\alpha-\beta^a\mathcal{D}_aK_{ij}-2K_{a(i}\mathcal{D}_{j)}\beta^a\right) .
\eea
In line with much of the literature we set the shift vector to zero.

The system now becomes
\beas
  ds^2&=&-\alpha^2dt^2+h_{ij}dx^idx^j, \\
  n^\mu&=&\left(\frac{1}{\alpha},\mathbf{0}\right), \quad n_\mu=(-\alpha,\mathbf{0}), \quad K_{ij}=-\frac{1}{2\alpha}\dot{h}_{ij}, \\
  \mathcal{R}+K^2-K^i_jK^j_i&=&16\pi G\rho+2\Lambda, \quad D_j\left(K^{ij}-h^{ij}K\right)=8\pi Gj^i, \\
  \frac{1}{\alpha}\dot{K}_{ij}&=&\mathcal{R}_{ij}-2K^a_iK_{aj}+KK_{ij}-8\pi GS_{ij}+4\pi Gh_{ij}\left(S-\rho\right)-\Lambda h_{ij}
   -\frac{1}{\alpha}\mathcal{D}_i\mathcal{D}_j\alpha .
\eeas

\section{Averaging in Cosmology}
\subsection{The Averaging Problem and Averaging Procedures}
\label{Averaging}
As we mentioned in section \ref{intro}, the Einstein equations have to be averaged over some domain to give an accurate description of the large scale behaviour of the universe. Since there is no physical meaning in the comparison of tensor quantities at different spacetime points, appropriate averaging is very involved. Despite many promising approaches such as Zalaletdinov's macroscopic description of gravity \cite{Zalaletdinov04} and the regional averaging and scaling approach by Buchert and Carfora \cite{BuchertCarfora02}, no procedure is yet agreed upon and work in this area is ongoing (e.g. \cite{Behrend07, Behrend08}).

Lacking this process at current, the standard approach employs the following. Let
\be
  V=\int_{\mathcal{D}}\sqrt{h}d^3\mathbf{x}
\ee
be the definition of the volume of a domain $\mathcal{D}\subset\Sigma$ smaller than the Hubble volume. Then one defines the average of a quantity $A$ in this domain as
\be
\label{AverageDef}
  \av{A}=\intdom{A} .
\ee
This process has several drawbacks. Most importantly it is not invariant under a change of coordinates. Therefore, the averaged quantities lose their tensor character and moreover depend on the coordinate system in which the averaging was performed. Nevertheless, it is believed that this averaging process does make sense for scalar quantities. See \cite{StoegerEtAl99} for a detailed discussion and an improved averaging process that produces quantities with approximate tensorial character in the weak field limit. To employ (\ref{AverageDef}) we will project all tensor quantities onto scalars.

\subsection{Average Hubble Rates, Scale Factors and Averaged Cosmological Equations}
\label{Averages}
We define an average Hubble rate by
\bdm
  3\frac{\dom{\dot{a}}}{\dom{a}}=\frac{\dot{V}}{V}=\frac{1}{V}\frac{\partial}{\partial t}\int_{\mathcal{D}}\sqrt{h}d^3\mathbf{x}
   =\intdom{\frac{1}{2}h^{ij}\dot{h}_{ij}} .
\edm
Using the extrinsic curvature we thus have
\be
  3\frac{\dom{\dot{a}}}{\dom{a}}=-\av{\alpha K};
\ee
that is, the effective Hubble rate in a domain is given by the averaged trace of the extrinsic curvature in that domain, weighted by the lapse function. We can also quickly find the commutator between time and space derivatives,
\be
\label{TimeAverage}
  \av{\dot{A}}=\frac{\partial}{\partial t}\av{A}+\frac{\dot{V}}{V}\av{A}+\av{A\alpha K}
   =\frac{\partial}{\partial t}\av{A}+3\frac{\dom{\dot{a}}}{\dom{a}}\av{A}+\av{A\alpha K} .
\ee

Consider first the Hamiltonian constraint; we can immediately see that
\bdm
  \av{\alpha^2\mathcal{R}}+\av{\left(\alpha K\right)^2}-\av{\alpha K^i_j\alpha K^j_i}=16\pi G\av{\alpha^2\rho}+2\Lambda\av{\alpha^2} .
\edm
This is quickly converted into
\be
\label{AvFriedmann}
  \left(\frac{\dom{\dot{a}}}{\dom{a}}\right)^2=\frac{8\pi G}{3}\av{\alpha^2\rho}+\frac{\Lambda}{3}\av{\alpha^2}
   -\frac{1}{6}\left(\Qd+\Rd\right)
\ee
with Buchert's kinematical backreaction
\be
  \Qd=\av{\alpha^2\left(K^2-K^i_jK^j_i\right)}-\frac{2}{3}\av{\alpha K}^2
\ee
and
\be
  \Rd=\av{\alpha^2\mathcal{R}}
\ee
a correction arising from the spatial curvature. Equation (\ref{AvFriedmann}) can be interpreted as an averaged Friedmann equation in the domain under consideration.

The trace of the evolution equation for the extrinsic curvature is,
\be
  \alpha\dot{K}-\alpha K^i_j\alpha K^j_i+\alpha\mathcal{D}^i\mathcal{D}_i\alpha=4\pi G\alpha^2(\rho+S)-\alpha^2\Lambda
\ee
which we have multiplied through by $\alpha^2$ for simplicity of manipulation. Now,
\bdm
  \av{\alpha\dot{K}}=\av{(\alpha K)}^\cdot+\av{\alpha^2K^2}-\av{\alpha K}^2-\av{\dot{\alpha}K},
\edm
and using the averaged Friedmann equation (\ref{AvFriedmann}) we find that the evolution equation averaged in the domain under consideration is
\be
\label{AvRay}
  \frac{\dom{\ddot{a}}}{\dom{a}}=-\frac{4\pi G}{3}\av{\alpha^2(\rho+S)}+\frac{\Lambda}{3}\av{\alpha^2}
   +\frac{1}{3}\left(\Qd+\Pd\right)
\ee
with
\be
  \Pd=\av{\alpha D^iD_i\alpha}-\av{\dot{\alpha}K},
\ee
Buchert's dynamical backreaction. We interpret (\ref{AvRay}) as an effective Raychaudhuri equation. Note that in the synchronous gauge, $\dom{\mathcal{P}}\equiv 0$.

Finally we have the so-called ``integrability condition'' which arises from the total continuity equation. We can find this by taking the time-derivative of the Friedmann equation and substituting in the Raychaudhuri and Friedmann equations. Writing $\dom{H}=\av{\mathcal{H}}$ where $\mathcal{H}=-\alpha K$ is a local Hubble rate, and using local matter continuity
\be
  \dot{\rho}+3\mathcal{H}\left(\rho+\frac{1}{3}\mathcal{S}\right)=0
\ee
we can ultimately find the integrability condition
\bea
\lefteqn{
  \frac{\partial_t\left(\dom{a}^6\dom{\mathcal{Q}}\right)}{\dom{a}^6}+\frac{\partial_t\left(\dom{a}^2\av{\alpha^2\mathcal{R}}\right)}{\dom{a}^2} 
   =12\av{\alpha^2\left(\frac{8\pi G}{3}\rho+\frac{\Lambda}{3}\right) \frac{\dot{\alpha}}{\alpha}} } \nonumber \\ &&+16\pi G\left(\av{\alpha^2\mathcal{S}}\av{\mathcal{H}}-\av{\alpha^2\mathcal{S}\mathcal{H}}\right)+6\Lambda\left(\av{\alpha^2\mathcal{H}}-\av{\alpha^2}\av{H}\right)-4\dom{H}\dom{\mathcal{P}}.
\eea
It is easy to see that in dust-filled synchronous gauge models, with $\alpha=1$ and $\mathcal{S}=0$, the source on the right-hand side vanishes.  For non-dust models, there is still an extra source term dependant on the local pressure and the local and averaged Hubble rates. In \cite{LiSchwarz07}, the authors make much use of the integrability condition, employing it iteratively to recover the backreaction at higher-orders in perturbation theory\footnote{See also \cite{LiSchwarz07-2} in which the same authors employ the integrability condition to extend their analysis to scales of 50Mpc; this paper was completed at the same time as the first version of the current work.}. Since the authors employ synchronous gauge and we employ Newtonian gauge our results are not directly comparable, but theirs is an interesting approach and exploits the simple nature of the synchronous gauge integrability condition to extend the analysis as far as perturbation theory holds.

\section{The Connection with Linear Perturbation Theory and the Size of the Backreaction}
\subsection{Backreaction Quantities in Newtonian Gauge}
\label{NewtGauge}
The Buchert equations are exact for any inhomogeneous model; for further progress we must specify this model. An increasingly common approach is to employ a model that averages out to be Robertson-Walker on large scales but is a relatively realistic approximation to the local universe on smaller scales (for a recent study in this area sharing some of our aims see for example \cite{KhosraviEtAl07}). We choose instead to use a linearly-perturbed FLRW model, employing Newtonian gauge. While much of the previous study in this field has been undertaken in synchronous gauge (see for example \cite{LiSchwarz07, LiSchwarz07-2}), this gauge contains unphysical modes and the metric perturbation can grow to be relatively large. While this is not necessarily an issue, Newtonian gauge is unambiguous and the variable $\phi$ remains small and well-defined across almost all scales. Moreover, it is easily incorporated into Boltzmann codes and should serve as a complementary probe to the previous studies, although it should be remembered that we cannot directly compare calculations performed in different gauges. As we will not employ the integrability condition, the complication in its source term is not for us an issue. It has been appreciated for some time \cite{Wetterich01,Rasanen03} that the impact on large-scale evolution from perturbations is not expected to be large, something clearly seen in Newtonian gauge; since the perturbations themselves are consistently small across almost all scales the impact from backreaction will be at most of the order of $10^{-5}$. To our knowledge, however, this has not been fully quantified. It would be interesting in particular to study the behaviour of the corrections to the standard FLRW model as a function of redshift.

We consider a universe filled with a cosmological constant and pressureless dust. This will serve as a reasonable approximation to the current universe on the largest scales. Working to first-order in the gravitational potential $\phi$, we consider only scalar perturbations and work with the line-element
\be
  ds^2=-(1+2\phi)dt^2+a^2(t)(1-2\phi)\delta_{ij}dx^idx^j
\ee
where the potentials are equal as a pure dust fluid implies vanishing anisotropic stresses, and the scale factor $a$ is distinct from the averaged scale factor $\dom{a}$. We neglect the tensor perturbations as sub-dominant to the scalars although a full approach should naturally take these into account. When performing our averages across scales approaching the Hubble horizon, we consider the quadratic order in linear perturbations and neglect the averages of pure first- and second-order perturbations. Were we to consider averaging across relatively small scales then these neglected terms should also be taken into account for a full evaluation. Likewise we can neglect the second-order vector and tensor perturbations that scalars inevitably source, but again these should be taken into account for a full treatment on small scales, a problem complicated by the lack of usable covariant averaging procedures. Since we are considering purely scalar perturbations, all vector quantities can be written as the gradient of a scalar quantity; in particular, $v^i=-\partial^i\psi$ for some velocity potential $\psi$.

Previous studies have occasionally extended their domain into super-horizon scales (e.g. \cite{GeshnizjaniBrandenberger02,GeshnizjaniBrandenberger03,MartineauBrandenberger05,MartineauBrandenberger05-2,KolbEtAl04,KolbEtAl04-2,KolbEtAl05}). This has been frequently criticised in the literature (e.g. \cite{LythMalikSasaki04,HirataSeljak05,Flanagan05,IshibashiWald06}). While it is perhaps still arguable that super-horizon isocurvature modes might generate an observable impact (e.g. \cite{GeshnizjaniBrandenberger03,KolbEtAl04-2,Parry06}), we will later restrict ourselves to adiabatic modes for which the effect is pure gauge and so choose our domain to be smaller than or equivalent to the horizon size.


By unambiguously identifying the 3+1 and Newtonian gauge co-ordinates and expanding quantities up to the second-order in perturbations (see also \cite{MukhanovEtAl90}), we can quickly see that
\be
  \alpha^2=1+2\phi, \quad \alpha=\sqrt{(1+2\phi)}\approx 1+\phi-\frac{1}{2}\phi^2, \quad
  \alpha^{-1}\approx 1-\phi+\frac{3}{2}\phi^2, \quad \alpha^{-2}\approx 1-2\phi+4\phi^2 .
\ee
We also get
\be
  h_{ij}=a^2(t)(1-2\phi)\delta_{ij}, \quad h^{ij}\approx a^{-2}(t)(1+2\phi+4\phi^2)\delta^{ij} .
\ee

We can now proceed to evaluate the geometric quantities we need in the averaged equations. The extrinsic curvature is
\be
  \alpha K^i_j=-\left(\frac{\dot{a}}{a}-\dot{\phi}(1+2\phi)\right)\delta^i_j .
\ee
Using the Ricci scalar on the spatial hypersurface the curvature correction is
\be
  \Rd=\av{\alpha^2\mathcal{R}}=\frac{2}{a^2}\av{\left(2\nabla^2\phi+3(\nabla\phi)^2+12\phi\nabla^2\phi\right)}
\ee
where $\nabla=(\partial/\partial x,\partial/\partial y,\partial/\partial z)$ as in standard vector calculus.

From the extrinsic curvature we can immediately see that
\bdm
  3\frac{\dom{\dot{a}}}{\dom{a}}=-\av{\alpha K}=3\av{\frac{\dot{a}}{a}-\dot{\phi}(1+2\phi)}
\edm
implying
\be
  \frac{\dom{\dot{a}}}{\dom{a}}=\frac{\dot{a}}{a}-\av{\dot{\phi}(1+2\phi)}
\ee
giving the relationship between the physical averaged scale factor and the scale factor employed in the perturbative approximation. If $\dot{\phi}=0$ then these coincide; this occurs in an Einstein de-Sitter universe, or when one considers a domain sufficiently small that its time variation can be neglected but sufficiently large that linear perturbation theory may be employed. (As an aside it should be commented in both these cases that feedback from the backreaction will rapidly break the condition $\dot{\phi}=0$ -- even if this is at a negligible level.)

We can now rapidly find the kinematical backreaction,
\be
  \Qd=6\left(\av{\left(\frac{\dot{a}}{a}-\dot{\phi}(1+2\phi)\right)^2}-\av{\frac{\dot{a}}{a}-\dot{\phi}(1+2\phi)}^2\right)
    =6\left(\av{\dot{\phi}^2}-\av{\dot{\phi}}^2\right) .
\ee

Using the covariant derivative on the 3-surface and
\be
  \dot{\alpha}K=3\dot{\phi}^2-3\frac{\dot{a}}{a}\dot{\phi}(1-2\phi),
\ee
the dynamical backreaction is
\be
  \Pd=3\frac{\dot{a}}{a}\av{\dot{\phi}(1-2\phi)}-3\av{\dot{\phi}^2}+\frac{1}{a^2}\av{\nabla^2\phi+2\phi\nabla^2\phi-2(\nabla\phi)^2} .
\ee
This gives us the geometric information that we need.

Consider now the fluids. We separated the stress-energy tensor with respect to the normal vector
\be
  n^\mu=\left(1-\phi+\frac{3}{2}\phi^2,\mathbf{0}\right) .
\ee
However, the stress-energy tensor employed in cosmological codes is defined with respect to the 4-velocity and for the case of pressureless dust is typically taken to be
\be
  T_{\mu\nu}=\varepsilon u_\mu u_\nu
\ee
where $\varepsilon$ is the energy density as defined in this reference frame, which will be linearised to $\varepsilon=\bkr(1+\delta)$ for an average $\bkr$. The density $\rho$ and FLRW density $\varepsilon$ thus do not coincide unless the velocity is orthogonal to the normal, which is not in general the case. The 4-velocity, to second-order in first-order perturbations, can be seen to be
\be
  u^\mu=\left(1-\phi+\frac{1}{2}\left(a^2v^2+3\phi^2\right),v^i\right), \quad
  u_\mu=\left(-\left(1+\phi+\frac{1}{2}\left(a^2v^2-\phi^2\right)\right),a^2(1-2\phi)v_i\right)
\ee
with $v_i=\delta_{ij}v^j$. Then the density is
\be
  \rho=T_{\mu\nu}n^\mu n^\nu=\varepsilon\left(u_\mu n^\mu\right)^2=\varepsilon\left(1+\frac{1}{2}a^2v^2\right)^2=\varepsilon\left(1+a^2v^2\right) .
\ee
Scaled by the lapse function, the quantity that enters the Buchert equations is then
\be
  \av{\alpha^2\rho}=\bkr\left(1+\delta+2\phi+a^2v^2+2\phi\delta\right)
\ee
and so we define a ``density correction''
\be
  \Td=\frac{8\pi G}{3}\bkr\av{\delta+2\phi+2\phi\delta+a^2v^2} .
\ee

The Friedmann equations can then be written as
\be
  \left(\frac{\dom{\dot{a}}}{\dom{a}}\right)^2=\frac{8\pi G}{3}\bkr_m+\frac{\Lambda}{3}+\Delta F, \quad
  \frac{\dom{\ddot{a}}}{\dom{a}}=-\frac{4\pi G}{3}\bkr_m+\frac{\Lambda}{3}+\Delta R
\ee
with corrections
\be
\label{DeltaFrieds}
  \Delta F=\Td-\frac{1}{6}\left(\Qd+\Rd\right), \quad
  \Delta R=-\frac{1}{2}\Td+\frac{1}{3}\left(\Qd+\Pd\right) .
\ee
The effective pressure and density of the corrections are then
\be
\label{EffectiveFluid}
  \frac{8\pi G}{3}\rho_{\mathrm{eff}}=\Td-\frac{1}{6}\left(\Qd+\Rd\right), \quad
  16\pi Gp_{\mathrm{eff}}=\frac{1}{3}\Rd-\Qd-\frac{4}{3}\Pd
\ee
and so the effective equation of state is
\be
\label{EoS}
  w_{\mathrm{eff}}=-\frac{1}{3}\left(\frac{\Rd-4\Pd-3\Qd}{\Rd-6\Td+\Qd}\right) .
\ee
Note that these definitions differ from those presented before due to the presence of the $\Td$ term, arising from directly comparing with a reference FLRW model.

We take the domain to be large enough to allow us to neglect the averages of first order quantities on the background. Specifically, we take the domain size to be of the order of the comoving Hubble scale, across which first-order averages can consistently be neglected. While this implies that our analysis can only be taken to apply on the very largest scales, taking the domain to be so large also implies that we can invoke the ergodic principle to convert our spatial averages into ensemble averages; we exploit this in the next section. On such large scales, the modifications become
\bea
\label{Rd}
  \Rd&=&\frac{6}{a^2}\av{(\nabla\phi)^2+4\phi\nabla^2\phi}, \\
\label{Pd}
  \Pd&=&6\av{\dot{\phi}^2}, \\
\label{Qd}
  \Qd&=&-6\frac{\dot{a}}{a}\av{\dot{\phi}\phi}-3\av{\dot{\phi}^2}+\frac{2}{a^2}\av{\phi\nabla^2\phi-(\nabla\phi)^2}, \\
\label{Td}
  \Td&=&\frac{8\pi G}{3}\bkr\av{2\phi\delta+a^2v^2} .
\eea
We can see that, depending on the signs of $\av{\phi\nabla^2\phi}$ and $\av{\phi\delta}$, for an Einstein de-Sitter universe with $\dot{\phi}=0$ (as considered in \cite{TanakaEtAl06}), one might get either an enhancement or reduction of both the effective Hubble rate and of the effective acceleration.

\subsection{Ergodic Averaging}
\label{ErgodicAveraging}
The simplest attack on the above formalism is to employ a Boltzmann code such as cmbeasy \cite{Doran03} or cmbfast \cite{Seljak96}. The use of linear theory automatically implies that we are working on relatively large scales; we can render the system developed in the last section more tractable by taking our domain to approach the Hubble volume itself -- large enough, at least, that we can employ the ergodic theorem and convert the spatial averages into averages across a statistical ensemble. (Similar approaches were also employed in \cite{Wetterich01,Rasanen03}.)

Consider first the general average $\av{A(t,\mathbf{x})B(t,\mathbf{x})}$ where $\{A,B\}\in\{\phi,\delta,v,\dot{\phi}\}$. Then
\be
  \av{A(t,\mathbf{x})B(t,\mathbf{x})}=\int_{\mathbf{k}}\int_{\mathbf{k}'}\av{A(t,\mathbf{k})B^*(t,\mathbf{k}')}e^{-i(\mathbf{k-k}')\cdot\mathbf{x}}\frac{d^3k}{(2\pi)^3}\frac{d^3k'}{(2\pi)^3} .
\ee
At linear order and in the absence of decoherent sources, different wavemodes are decoupled from one another and the evolution equations depend only on the magnitude of $k$ and so we can write
\be
  A(t,\mathbf{k})=\alpha(\mathbf{k})A(t,k)
\ee
for any quantity. The initial correlation between two arbitrary quantities is
\be
  \av{\alpha(\mathbf{k})\beta^*(\mathbf{k}')}=\frac{2\pi^2}{k^3}\mathcal{P}_{\psi}(k)\delta(\mathbf{k-k}')
\ee
where $\mathcal{P}_\psi(k)$ is the primordial power spectrum of the metric fluctuations. Inserting this into the correlation and integrating once yields
\be
  \av{A(t,\mathbf{x})B^*(t,\mathbf{x})}\approx\int_k\mathcal{P}_\psi(k)A(t,k)B^*(t,k)\frac{dk}{k}.
\ee

The averages involving gradients are only a little more complicated; for $\av{(\nabla\phi)^2}$ we have
\bea
  \av{(\nabla\phi)^2}&=&\int_{\mathbf{k}}\int_{\mathbf{k}'}(-i\mathbf{k})(i\mathbf{k}')\av{\phi(t,\mathbf{k})\phi^*(t,\mathbf{k}')}e^{-i(\mathbf{k-k}')\cdot\mathbf{x}}\frac{d^3k}{(2\pi)^3}\frac{d^3k'}{(2\pi)^3} \\
  &=&\int_kk^2\mathcal{P}_\psi(k)\left|\phi(t,k)\right|^2\frac{dk}{k} ,
\eea
which is positive-definite as it should be. The other curvature term is
\bea
  \av{\phi\nabla^2\phi}&=&-\int_{\mathbf{k}}\int_{\mathbf{k}'}\mathbf{k}'^2\av{\phi(t,\mathbf{k})\phi^*(t,\mathbf{k}')}e^{-i(\mathbf{k-k}')\cdot\mathbf{x}}\frac{d^3k}{(2\pi)^3}\frac{d^3k'}{(2\pi)^3} \\
  &=&-\int_kk^2\mathcal{P}_\psi(k)\left|\phi(t,k)\right|^2\frac{dk}{k} .
\eea
This term is then negative definite.

The kinematical, dynamical, curvature and density corrections (\ref{Rd}-\ref{Td}), then become
\bea
  \Rd&=&-\frac{18}{a^2}\int k^2\mathcal{P}_\psi(k)\left|\phi(t,k)\right|^2\frac{dk}{k}, \\
  \Qd&=&6\int\mathcal{P}_\psi(k)\left|\dot{\phi}(t,k)\right|^2\frac{dk}{k}, \\
  \Pd&=&-3\int\mathcal{P}_\psi(k)\left(\left|\dot{\phi}(t,k)\right|^2+\frac{\dot{a}}{a}\left(\phi(t,k)\dot{\phi}^{*}(t,k)+\phi^*(t,k)\dot{\phi}(t,k)\right)+\frac{4}{3}\frac{k^2}{a^2}\left|\phi(t,k)\right|^2\right)\frac{dk}{k}, \\
  \Td&=&\frac{8\pi G}{3}\bkr\left(\int k^2\mathcal{P}_\psi(k)a^2(t)\left|v(t,k)\right|^2\frac{dk}{k}-\int\mathcal{P}_\psi(k)\left(\phi(t,k)\delta^*(t,k)+\phi^*(t,k)\delta(t,k)\right)\frac{dk}{k}\right) .
\eea
For a more realistic approach where we separate the baryons from the cold dark matter one can readily see that the contribution from the density can be expanded as
\bea
  \Td&=&\frac{8\pi G}{3}\int k^2\mathcal{P}_\psi(k)\left(\bkr_b\left|v_b(t,k)\right|^2+\bkr_c\left|v_c(t,k)\right|^2\right)\frac{dk}{k}
  \nonumber \\ && \quad
   -\int\mathcal{P}_\psi(k)\left(\phi(t,k)\left(\bkr_b\delta_b^*(t,k)+\bkr_c\delta_c^*(t,k)\right)+\phi^*(t,k)\left(\bkr_b\delta_b(t,k)+\bkr_c\delta_c(t,k)\right)\right)\frac{dk}{k} .
\eea

Since Boltzmann codes tend to be written in conformal time we must these into conformal time expressions using $d/dt=(1/a)d/d\eta$; denoting conformal time quantities with a tilde we thus have
\be
  v=\frac{\tilde{v}}{a}, \quad
  \dot{\phi}=\frac{\phi'}{a}, \quad
  \frac{\dot{a}}{a}=\frac{1}{a}\frac{a'}{a}
\ee
in the expressions above.

\section{Results}
\subsection{Corrections to the FLRW Picture from Linear Perturbations}
\label{Backfast}
We incorporate the above formalism into the cmbeasy Boltzmann code and run consistency checks with the cmbfast code, modified to output conformal Newtonian quantities. We will evaluate the terms across all post-recombination redshifts and imploy and infra-red cut-off at the comoving Hubble scale, $k_{\mathrm{min}}=1/\eta$, avoiding the unphysical gauge-dependent super-horizon contributions. The small-scale limit $k_{\mathrm{max}}$ of our domain, determined from the stability of the integration with respect to changing $k_{\mathrm{max}}$ is about $k_{\mathrm{max}}\approx 30$MPc$^{-1}$ and we integrate to $k_\mathrm{max}=100$Mpc$^{-1}$. Naturally, we do not claim that our results at such small scales are complete, merely that we are evaluating the contribution of such large-scale modes on these scales.

At a linear level, there are scaling relations that hold to a high level of accuracy; it is obvious from Poisson's equation that
\be
  \phi\propto\delta/k^2 ;
\ee
from Euler's equation $\dot{v}+(\dot{a}/a)v\propto\nabla\phi$ we can also predict that
\be
  \left|v\right|\propto\delta/k .
\ee
From here, one may immediately state that on smaller scales where $1/k^4\ll 1/k^2$ and for an approximately scale-invariant primordial power spectrum,
\be
  \Qd(k)\propto\frac{\delta^2}{k^4}, \quad \Td(k)\propto\Pd(k)\propto\Rd(k)\propto\frac{\delta^2}{k^2}
\ee
where $\Qd(k)$ is the integrand of $\Qd$. $\Qd$ is thus generally subdominant to the other corrections except on very large scales. This also implies that each correction is approximately of the form
\be
  \mathcal{A}_D=\alpha\int\left|\delta\right|^2\frac{\mathcal{P}_\psi}{k^3}dk
\ee
for some constant $\alpha$, reminiscent of the approximation in equation (32) of \cite{Wetterich01}.

\begin{center}
\begin{figure}
\includegraphics{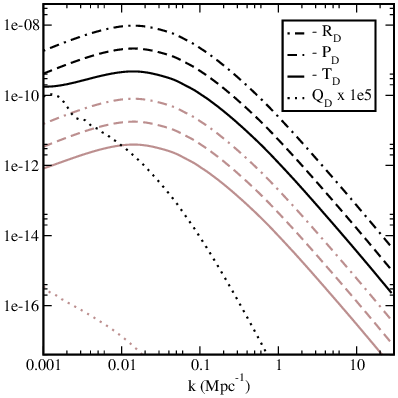}\quad\includegraphics{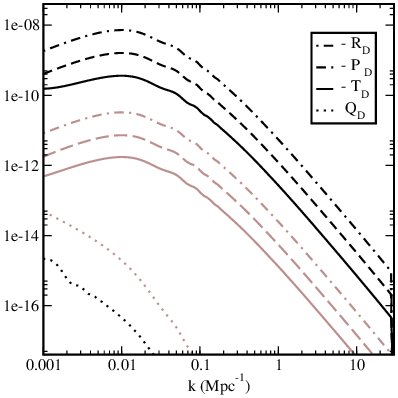}
\caption{The correction terms $\Td$, $\Pd$, $\Rd$ and $\Qd$ as a function of $k$ at $z=10$ (black) and $z=0$ (grey) for (left) the sample Einstein de-Sitter model and (right) the WMAPIII concordance model.}
\label{Figure-EdSComparisonz10z0}
\end{figure}
\end{center}

\begin{center}
\begin{figure}
\includegraphics{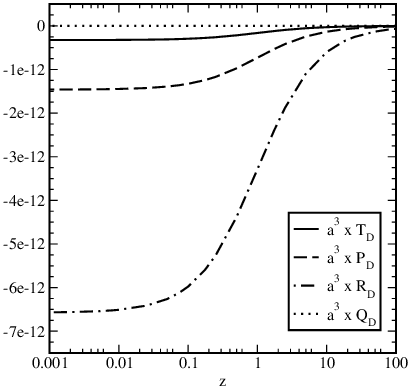}\quad\includegraphics{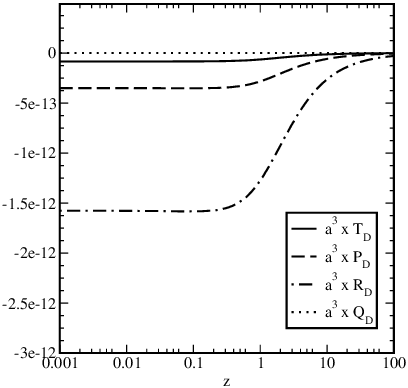}
\caption{The evolution of the corrections for (left) the sample EdS model and (right) the WMAPIII concordance model.}
\label{Figure-EdSEvolution}
\label{Figure-LCDMEvolution}
\end{figure}
\end{center}

For our models we take a low-Hubble constant Einstein de-Sitter model with $\Omega_b=0.05$, $\Omega_m=1$ and $h=0.41$ and a WMAPIII $\Lambda$CDM concordance model. The left panel in Figure \ref{Figure-EdSComparisonz10z0} shows the (integrands of the) four different correction terms at redshifts of $z=10$ and $z=0$ for the EdS case and the right panel the same for the $\Lambda$CDM case, and in Figure \ref{Figure-EdSEvolution} we present the correction terms as a function of $z$ for the EdS and $\Lambda$CDM cases; the premultiplication by $a^3(z)$ acts as a volume normalisation. The subdominance of $\Qd(k)$ and proportionality between $\Rd$, $\Td$ and $\Pd$ is very clear, with $\Rd$ the strongest correction.

\begin{center}
\begin{figure}
\includegraphics{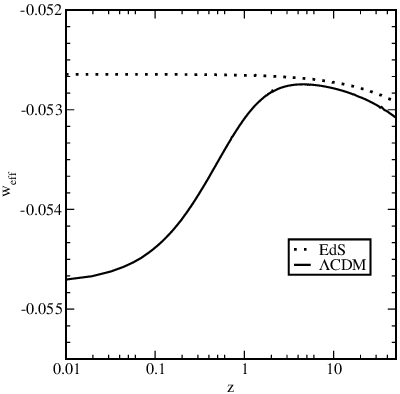}\quad\includegraphics{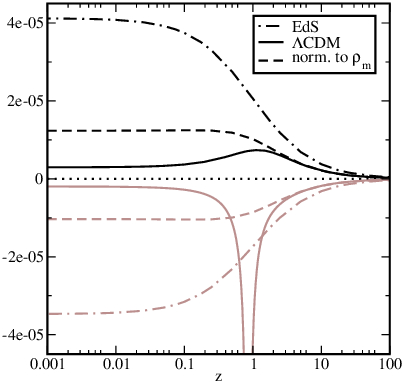}
\caption{Left: The effective equation of state $w_\mathrm{eff}$ as a function of redshift for a sample Einstein de-Sitter and WMAP concordance models. Right: The impact of the corrections onto the Friedmann equation (black) and the Raychaudhuri equation (brown), normalised in the $\Lambda$CDM case to the matter content (dashed) and to the standard equations (solid).}
\label{Figure-weff}
\label{Figure-Impacts}
\end{figure}
\end{center}

From the proportionality of the other corrections (which holds up to relatively large scales) we can write $\Td=t\Rd$ and $\Pd=p\Rd$ and express the equation of state (\ref{EoS}) as
\be
  w_{\mathrm{eff}}\approx-\frac{1}{3}\left(\frac{1-4p}{1-6t}\right) .
\ee
which is simple to evaluate numerically by selecting some pivot scale on which to evaluate the ratios $p(z)$ and $t(z)$. Doing so at the current epoch and a pivot scale of $k=0.01$Mpc$^{-1}$, we find $p\approx 2/9$ and $t\approx 1/20$; both of these are also only slowly evolving. This gives an estimate of the average equation of state as $w_\mathrm{eff}\approx -1/19$. In the left panel of Figure \ref{Figure-weff} we plot the evolution of $w_\mathrm{eff}$ for both the model EdS and $\Lambda$CDM cases evaluated directly from equation (\ref{EoS}). In both cases the effective equation of state from linear perturbations remains around the order $w_\mathrm{eff}\approx -1/19$ and the correction terms to the usual FLRW as a whole thus act as a form of non-standard dark matter and \emph{not} as a dark energy. These results compare reasonably well with the estimate in \cite{Wetterich01} where the author found $w_\mathrm{eff}\approx-1/27$\footnote{They also found $w_\mathrm{eff}\approx -1/15$ for a clustering cosmon field but we have not considered such a component here.}.

To quantify the impact on the Friedmann and Raychaudhuri equations, define
\be
  F_m=\frac{8\pi G\bkr_{m0}}{3a^3}, \quad R_m=\frac{4\pi G\bkr_{m0}}{3a^3}
\ee
as the standard contribution from baryons and CDM. We can then consider $\Delta F/F_m$ and $\Delta R/R_m$ as a well-defined measure of the impact of the correction terms in both EdS and $\Lambda$CDM models. By equation (\ref{EffectiveFluid}), $\Delta F/F_m$ for an EdS model is just $\rho_\mathrm{eff}/\bkr_m$. For the sake of clarity we choose to focus on deviations from the standard behaviour but, naturally, we can re-express these quantities in terms of the effective energy density and pressure.

For the EdS case we can see in Figure \ref{Figure-Impacts} that the total impact from linear perturbations tends to $\sim 4\times 10^{-5}$ on the Friedmann equation and $\sim -3.2\times 10^{-5}$ on the Raychaudhuri equation; the corrections thus act with a positive effective density and with insufficiently negative pressure to accelerate the universe and instead act to decelerate it at a negligible level. The dashed curves in Figure \ref{Figure-Impacts} are for the $\Lambda$CDM case; they tend to impacts of $\sim 1.3\times 10^{-5}$ on the Friedmann equation and $\sim -1\times 10^{-5}$ on the Raychaudhuri equation. The behaviour, as might be expected, is qualitatively similar to that in the EdS case and the impact is significantly less (by a factor of roughly $10/3$ at the current epoch).

This compares reasonably well with the approximation in \cite{Wetterich01} which in our notation is
\be
  \frac{\rho_\mathrm{eff}}{\overline{\rho}_m}=\frac{9a^2}{4}(8\pi G\overline{\rho}_m)\int\frac{\mathcal{P}_\psi}{k^3}dk .
\ee
Evaluating this approximation for the EdS universe, we find that this [under/over]estimates the correction terms with respect to our more detailed study by a factor of about 30\%. In the same paper, the author found a ``cosmic virial theorem'',
\be
  \overline{p}_m=-p_\mathrm{eff}
\ee
where $\overline{p}_m$ is the correction to the matter pressure arising from gravitational interactions. Crudely modelling the matter pressure as $\overline{p}_m\approx (1/3)\overline{\rho}_m\av{v^2}$ we find that $\overline{p}_m$ underestimates $-p_\mathrm{eff}$, again by $\sim$30\%, but the proportionality between them holds remarkably well for a wide range of redshifts.

For the $\Lambda$CDM case we can consider an alternative normalisation that directly quantifies the impact on a universe with a cosmological constant,
\bdm
  \frac{\Delta F}{\dom{\dot{a}}/\dom{a}}=\frac{\Delta F}{F_m+\Lambda/3}, \quad
  \frac{\Delta R}{|\dom{\ddot{a}}/\dom{a}|}=\frac{\Delta R}{|R_m-\Lambda/3|} .
\edm
The benefit of doing so is that it demonstrates the declining contribution of the corrections with respect to the cosmological constant; however, it also introduces a singularity at a redshift of $z\approx 0.8$ when the reference FLRW universe undergoes a transition from deceleration to acceleration. The impact on both the Friedmann and Raychaudhuri equations tends to $\sim 10^{-6}$ and the maximum contribution is at a redshift of $z\approx 1.3$.

\subsection{Corrections to the FLRW Picture from Quasilinear Perturbations}
\label{Halofit}
In the previous section we evaluated the impact on the large-scale evolution of the universe from linear perturbations, demonstrating that it is small, as should be expected, but also that it acts as a dark matter and not a dark energy. In this section we employ the halo model to estimate the impact of perturbations on smaller, quasilinear scales which are much less understood.

The publicly-available Halofit code \cite{SmithEtAl02} converts a linear CDM power spectrum into the CDM power spectrum from the Halo model. We employ a modified Halofit code that instead takes the linear matter power spectrum and estimates the nonlinear matter power spectrum. The square-root of this power spectrum can then be used to estimate $\delta(k)$ and we can then recover the other relevant quantities from the scaling relations $v\propto\delta/k$, $\phi\propto\delta/k^2$. If we define
\be
  f(k)=\frac{\delta_{\mathrm{L}}(k)}{\delta_{\mathrm{NL}}(k)}
\ee
we can employ this as a scaling factor to recover estimates for the quasilinear behaviour of $\delta(k)$, $v(k)$, $\phi(k)$ and $\dot{\phi}(k)$. We can then employ the formalism we developed for the linear case to estimate the impact in the quasilinear case, retaining the same domain size. We should immediately note two caveats. Firstly, while the scaling relation for the velocity is extremely precise for linear perturbations it is not for non-linear perturbations and so we have automatically introduced a source of error. Perhaps more importantly, we are employing a formalism developed for linear perturbations in which different wavemodes decouple from one another, allowing us to separate the statistical problem into transfer functions and a primordial power spectrum. This does not hold for a nonlinear problem. For both of these reasons, our results are only intended to be taken as good approximations on quasilinear scales until the velocity virialises, at which point a more detailed study is necessary.

\begin{center}
\begin{figure}
\includegraphics{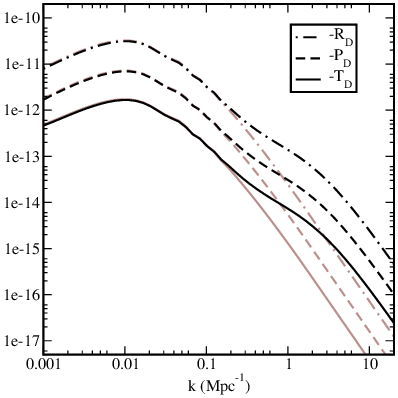}\quad\includegraphics{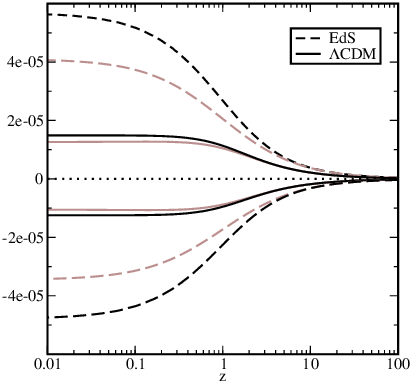}
\caption{Left: The halo model approximations to the quasilinear corrections for the WMAPIII concordance model at $z=0$. Right: The halo model approximations to the quasilinear corrections to the Friedmann (positive) and Raychaudhuri (negative) equations; the linear prediction is in grey.}
\label{Figure-NLLCDM}
\label{Figure-NLImpact}
\end{figure}
\end{center}

In Figure \ref{Figure-NLLCDM} we present approximations for the corrections to the $\Lambda$CDM model at redshift 0. As before, the kinematical backreaction is strictly negligible and as we have merely scaled the previous corrections by the same quantity, the effective equation of state remains unchanged. The total impacts on the Friedmann and Raychaudhuri equations for both are shown in Figure \ref{Figure-NLImpact}. We see that  although the halo model provides a boost in power on smaller scales, and even though $\Rd$ and $\Td$ in particular include factors of $k^2$ in the integral that increase the contribution from smaller scales, the impact is not significantly greater than from linear scales. More quantitively, for an EdS-universe the impact at the current epoch on the Friedmann equation is of order $5.6\times10^{-5}$ and on the Raychaudhuri equation is of order $-4.7\times10^{-5}$; normalised to the matter content, the impact at the current epoch from $\Lambda$CDM perturbations is of order $1.6\times10^{-5}$  on the Friedmann equation and of order $-1.4\times10^{-5}$ on the Raychaudhuri equation, normalised to the full evolution $4.4\times10^{-6}$ and $-2.7\times10^{-6}$, respectively.

\section{Discussion and Conclusions}
We have presented quantitive estimates of the corrections in Newtonian gauge to a standard perturbative FLRW model from an explicit averaging procedure, which can be separated into four distinct quantities: the kinematic backreaction, which remains strictly negligible across the scales of interest, the dynamic backreaction, the impact of spatial curvature and a correction to the FLRW energy density. In line with expectation, the impact from linear perturbations is insignificant and of the order of $10^{-5}$ for both $\Lambda$CDM and an Einstein de-Sitter model. Specifically, the effective energy density arising at the background level from inhomogeneities is $\rho_\mathrm{eff}\approx (4\times 10^{-5})\bkr_m$ and $\rho_\mathrm{eff}\approx(1.3\times 10^{-5})\bkr_m$ for the EdS and $\Lambda$CDM models respectively. Moreover, we have presented estimates arising from halo model corrections on quasilinear scales which are not much larger than those from the linear perturbations, the largest contribution arising from the quasilinear modes in an EdS universe remaining below $10^{-4}$; we find $\rho_\mathrm{eff}\approx(5.6\times 10^{-5})\bkr_m$ for the EdS model and $\rho_\mathrm{eff}\approx(1.6\times 10^{-5})\bkr_m$ for $\Lambda$CDM. This is in broad agreement with recent calculations from Vanderveld \emph{et. al.} estimating the total possible backreaction from observations of Type Ia supernovae and agrees with a growing consensus that the impact of backreaction across all scales remains at the level of $10^{-5}-10^{-4}$. Additionally, the total effective equation of state arising from the different modifications does not act as a dark energy and instead as a dark matter with a slowly-varying equation of state $w_\mathrm{eff}\approx -1/19$ for $z\in(0,10)$. The impact is to decelerate the universe and, contrary to other studies, the corrections impact positively on the Friedmann equation implying a positive effective pressure. (The discrepancy arises due to our definitions of the effective energy density and pressure of the corrections taking into account the differences in definition of energy density in an FLRW frame and with respect to our foliation.)

This does not, however, necessarily imply that there are no observational consequences arising from inhomogeneities, particularly on small scales, as our analysis is limited to the impacts in very large volumes and breaks down when mode-coupling and virialisation become significant. In particular, one can readily imagine situations in which the local Hubble rate is significantly larger than the global average (the so-called ``Hubble bubble''). The impact of local inhomogeneities directly on luminosity distances has also long been well studied and remains an active area of interest. Moreover, the issue of averaging in cosmology and general relativity in particular remains very much an active field. The average we have employed in this study is certainly not perfect; even were we to extend our approach to second-order perturbations it cannot cope with the vector and tensor perturbations that necessarily become significant.
Improving the average and applying it to nonlinear situations is an immediate aim.

The prospect for future study remains large. Our model here is necessarily simple and there are many avenues both analytical and numerical that can be pursued. Much interest is currently focusing on the exact swiss-cheese models and particularly on the impact on direct observables from these models. The prospect of directly calculating both the direct backreaction in different domain sizes and the impact on null geodesics (and hence luminosity distances) in different matter configurations from numerical simulations is also an exciting one. More immediately, one can imagine employing second-order perturbation theory in which the ambiguity between averages taken across the assumed zeroth-order manifold and those across the linearly-perturbed manifold may become significant; depending on the surface one defines a vanishing $\langle\phi_{(2)}\rangle$ across, one would expect a non-vanishing contribution on the other. In particular, if one defined second-order perturbations to vanish on average on the background then there would be a residual contribution to the linear perturbations even before any other backreaction is taken into account. This is an interesting issue that deserves more study.  More ambitiously, the fully non-linear approaches of Vernizzi and Langlois \cite{LangloisVernizzi05_1,LangloisVernizzi05_2,LangloisVernizzi06_1} and Enqvist \emph{et. al.} \cite{EnqvistEtAl06} might provide an interesting tool for studying the impacts of averaged nonlinear perturbations. The direct backreaction from more complicated systems than our linearly-perturbed FLRW universe might also be considerable; an immediate example, first considered in \cite{Wetterich01}, would be the backreaction from inhomogeneities in a cosmological scalar field which were shown to potentially have an impact of order unity. Such systems might also be considered in a fully non-linear manner \cite{LangloisVernizzi06_2}.

In summary, one may comment that backreaction and the impact of inhomogeneities are direct physical effects and should certainly be taken into account in any full treatment of the observables. However, the direct impact when evaluated for the concordance cosmology on very large scales is too small to provide an alternative to the dark energy.

\begin{acknowledgments}
The authors would like to thank Christof Wetterich, Michael Doran, Thomas Buchert, David Parkinson, David Mota, Thomas Dent and Otto Nachtmann for useful discussions and Christof Wetterich for comments on the script.
\end{acknowledgments}

\bibliography{paper}

\end{document}